\documentclass[manuscript]{emulateapj}

\shorttitle{Oscillating light wall above a sunspot light bridge}

\shortauthors{Yang et al.}

\journalinfo{Accepted for publication in ApJL}

\submitted{Accepted for publication in ApJL}

\begin{document}

\title{Oscillating light wall above a sunspot light bridge}

\author{Shuhong Yang\altaffilmark{1}, Jun Zhang\altaffilmark{1}, Fayu Jiang\altaffilmark{1}, and
Yongyuan Xiang\altaffilmark{2}}

\altaffiltext{1}{Key Laboratory of Solar Activity, National
Astronomical Observatories, Chinese Academy of Sciences, Beijing
100012, China; shuhongyang@nao.cas.cn}

\altaffiltext{2}{Fuxian Solar Observatory, Yunnan Observatories,
Chinese Academy of Sciences, Kunming 650011, China}

\begin{abstract}

With the high tempo-spatial \emph{Interface Region Imaging
Spectrograph} 1330 {\AA} images, we find that many bright structures
are rooted in the light bridge of NOAA 12192, forming a \emph{light wall}.
The light wall is brighter than the surrounding areas, and the wall
top is much brighter than the wall body. The New Vacuum Solar
Telescope H$\alpha$ and the \emph{Solar Dynamics Observatory} 171
{\AA} and 131 {\AA} images are also used to study the light wall
properties. In 1330 {\AA}, 171 {\AA}, and 131 {\AA}, the top of the
wall has a higher emission, while in the H$\alpha$ line, the wall
top emission is very low. The wall body corresponds to bright areas
in 1330 {\AA} and dark areas in the other lines. The top of the
light wall moves upward and downward successively, performing
oscillations in height. The deprojected mean height, amplitude,
oscillation velocity, and the dominant period are determined to be
3.6 Mm, 0.9 Mm, 15.4 km s$^{-1}$, and 3.9 min, respectively. We
interpret the oscillations of the light wall as the leakage of
\emph{p}-modes from below the photosphere. The constant brightness
enhancement of the wall top implies the existence of some kind of
atmospheric heating, e.g., via the persistent small-scale
reconnection or the magneto-acoustic waves. In another series of
1330 {\AA} images, we find that the wall top in the upward motion
phase is significantly brighter than in the downward phase. This
kind of oscillations may be powered by the energy released due to
intermittent impulsive magnetic reconnection.

\end{abstract}


\section{INTRODUCTION}

In the sunspot umbra, the overturning motion of the plasma is
hindered by the strong magnetic field, thus leading to a lower
temperature in the photosphere due to the reduced energy input
(Gough \& Tayler 1966). Within the umbra, light bridges (LBs) are
one kind of bright structures with incompletely suppressed
convection (Sobotka et al. 1993; Borrero \& Ichimoto 2011). LBs are
considered to have a deep anchoring in the convection zone (Lagg et
al. 2014). Faint LBs penetrate into the umbra, and strong LBs can
divide the umbra into separate umbral cores (Sobotka et al. 1994;
Rimmele 2008). The widths of LBs mainly range from no more than one
arcsecond to several arcseconds, and the typical brightness is
comparable to the penumbra. In LBs, the magnetic field is weaker and
the magnetic inclination is higher than in the neighboring umbra,
forming a magnetic canopy (Lites et al. 1991; R\"{u}edi et al. 1995;
Leka 1997; Jur{\v c}{\'a}k et al. 2006; Sobotka et al. 2013). The
umbral cores separated by a LB have either the opposite or the same
magnetic polarities (Zirin \& Wang 1990; Sobotka et al. 1994).

Sunspot oscillations have been extensively observed and investigated
for nearly half a century since their discovery (Beckers \& Tallant
1969). The oscillation power in a sunspot is non-uniformly
distributed. The oscillations with a period of about 3 min are
prominent in the umbra and the 5-min oscillations dominate the
penumbra (Christopoulou et al. 2000; Yuan et al. 2014a). In the
umbra, 5-min oscillations are greatly suppressed (Zirin \& Stein
1972; Stangalini et al. 2012). Umbral oscillations are explained as
standing slow mode magneto-acoustic waves between the photosphere
and transition region (Botha et al. 2011). The 3-min oscillations
appear as slow magneto-acoustic waves propagating along the magnetic
field lines from the umbra to the corona (de Moortel 2009; Kiddie et
al. 2012; Tian et al. 2014; Cho et al. 2015). The running penumbral
waves are observed to be disturbances propagating outward along the
penumbra with the period around 300 s (Zirin \& Stein 1972;
Tziotziou et al. 2006). The oscillations in LBs have not been well
studied until now. The atmosphere above LBs is often manifested in
forms of brightnings, jets, and surges (Shimizu et al. 2009).
Sobotka et al. (2013) analyzed the oscillations above a pore lacking
a penumbra, and found that the power dominant in the LB is around 4
mHz, i.e., 4.2 min. Recently, Yuan et al. (2014b) studied the
variations of LB intensity in solar active region (AR) 11836, and
their results showed that 5-min oscillations are significantly
presented in LBs and the 3-min oscillations in LBs are suppressed.

NOAA 12192 which occurred in the quite weak solar cycle 24 is the
largest AR since 1990 November (Thalmann et al. 2015; Sun et al.
2015). During the passage of AR 12192 across the visible solar disk
in 2014 October, the \emph{Interface Region Imaging Spectrograph}
(\emph{IRIS}; De Pontieu et al. 2014) was operated to observe the AR
with a field-of-view (FOV) of 119{\arcsec}$\times$119{\arcsec} and
obtained quite a lot of excellent data. Especially, a period of
about 1 hr around 06:00 UT on October 25 was spent observing the
light bridge dynamics as designed, and a series of high
tempo-spatial images were obtained. In the present study, we report
the discovery of oscillating light wall above a light bridge in AR
12192, using the jointed observations from the \emph{IRIS}, the
\emph{Solar Dynamics Observatory} (\emph{SDO}; Pesnell et al. 2012),
and the New Vacuum Solar Telescope (NVST; Liu et al. 2014) of the
\emph{Fuxian Solar Observatory} in China.

\begin{figure*}
\centering
\includegraphics
[bb=100 116 480 730,clip,angle=0,width=0.64\textwidth] {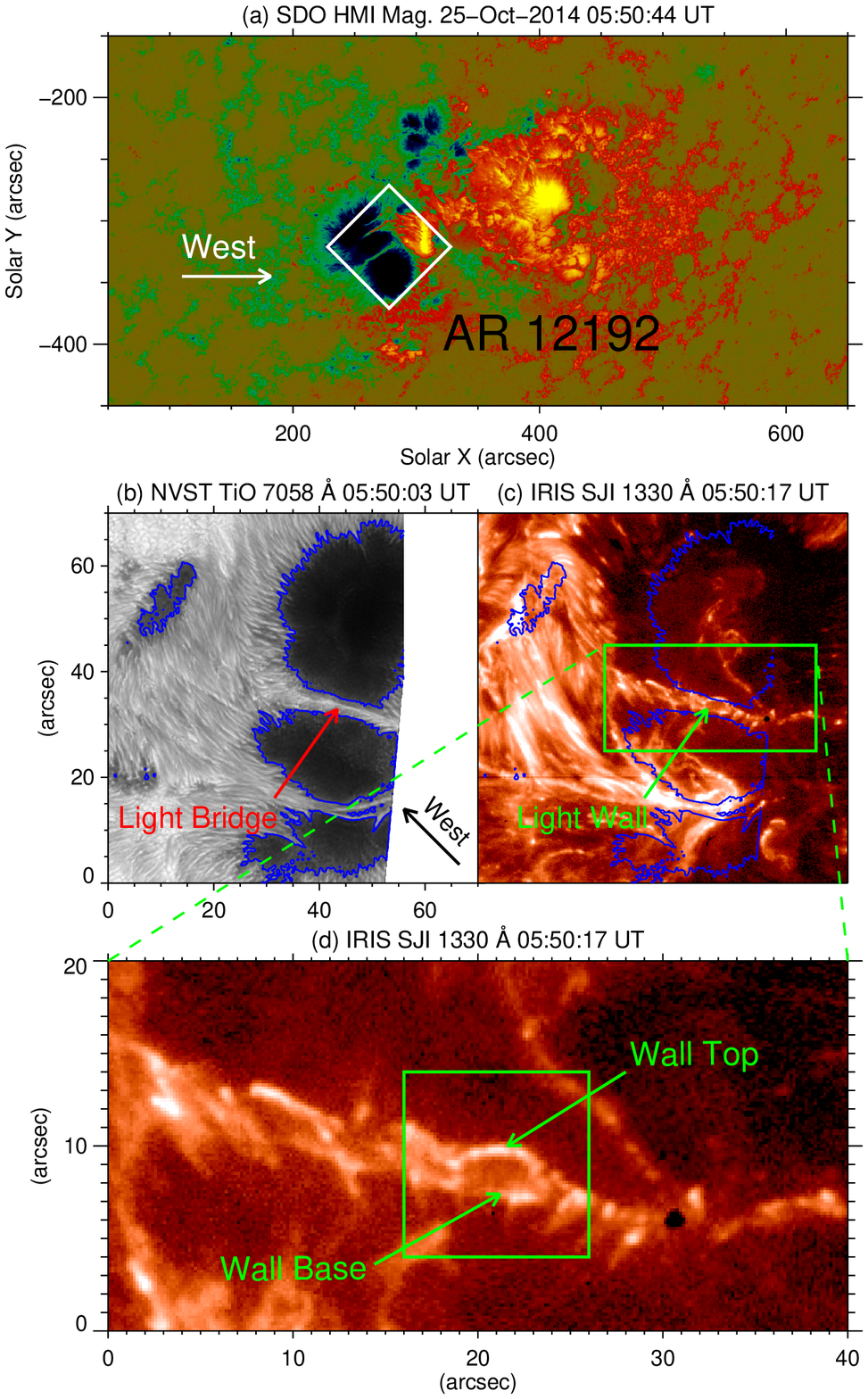}
\caption{Panel (a): \emph{SDO} HMI line-of-sight magnetogram
displaying the overview of AR 12192. Panels (b)-(c): NVST TiO 7058
{\AA} and \emph{IRIS} SJI 1330 {\AA} images (also see animation 1)
showing the light bridge and the light wall in the FOV outlined by
the square in panel (a). The blue curves in panels (b) and (c) are
the contours of the sunspot umbra. Panel (d): expanded view of the
1330 {\AA} image outlined by the rectangle in panel (c). The square
outlines the FOV of Figures 2(a)-(d). \label{fig}}
\end{figure*}

\begin{figure*}
\centering
\includegraphics
[bb=120 155 457 661,clip,angle=0,width=0.6\textwidth] {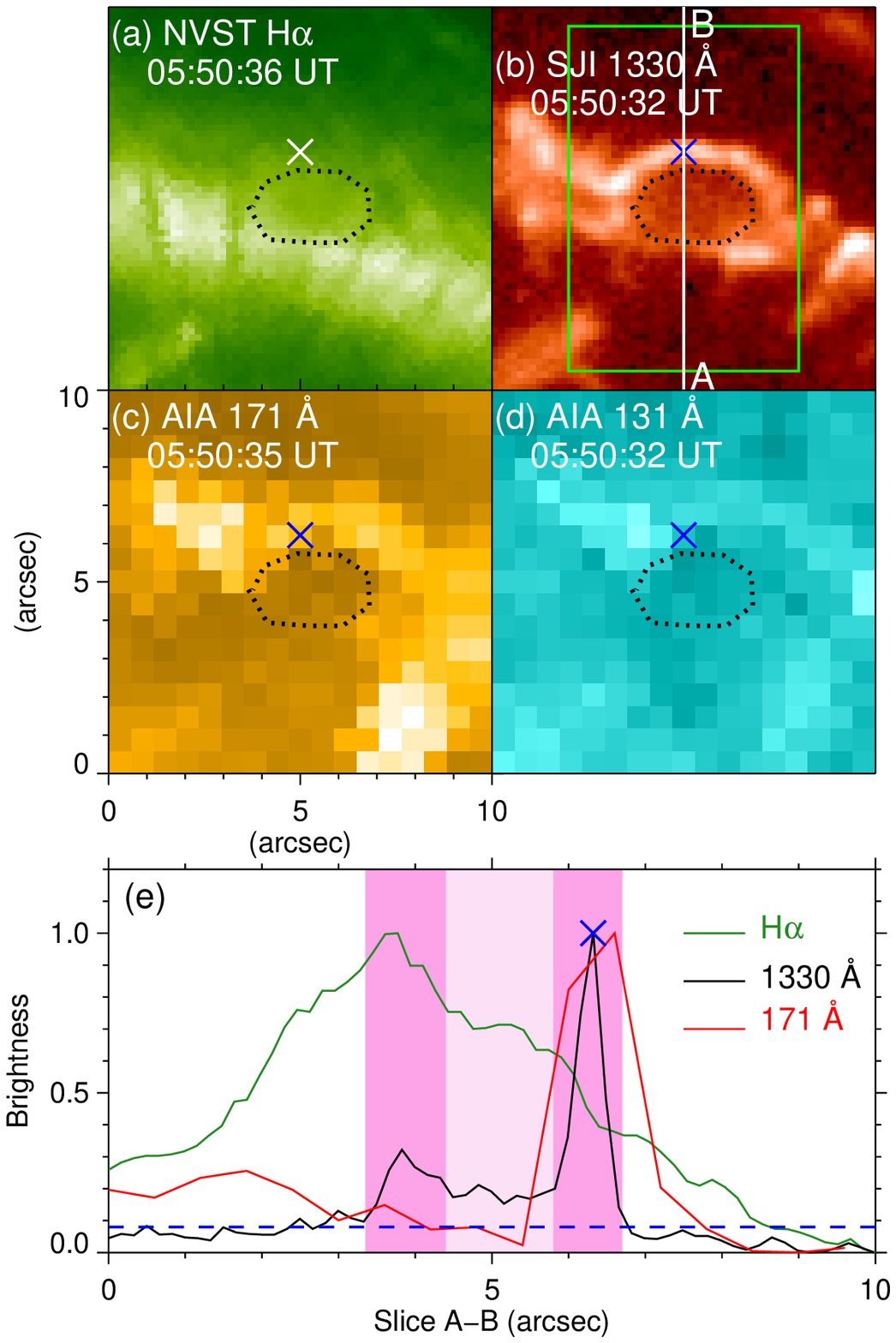}
\caption{ Panels (a)-(d): NVST H$\alpha$, SJI 1330 {\AA}, AIA 171
{\AA} and 131 {\AA} images showing the appearance of the light wall
at different temperatures. The dotted curves delineate the general
shape of a bubble-like structure identified in the 1330 {\AA} image.
The rectangle in panel (b) outlines the FOV of Figures 3(a1)-(a5).
Panel (e): multi-wavelength brightness along slice ``A-B" marked in
panel (b). The deep, light, and deep pink shadows mark the wall
base, wall body, and wall top, respectively. \label{fig}}
\end{figure*}

\section{OBSERVATIONS AND DATA ANALYSIS}

Two series of \emph{IRIS} slit-jaw 1330 {\AA} images (SJIs) are
adopted. The 1330 {\AA} passband contains emission from the strong C
II 1334/1335 {\AA} lines formed in the chromosphere and lower
transition region, and also includes continuum emission formed
around the temperature minimum (upper photosphere and lower
chromosphere). On October 25, the SJIs in 1330 {\AA} were taken from
05:17:13 UT to 06:30:18 UT with a spatial resolution of
0.{\arcsec}166 pixel$^{-1}$ and a cadence of 7 s. On October 29, the
1330 {\AA} SJIs were obtained from 15:30:00 UT to 18:17:49 UT with a
cadence of 16 s and a pixel size of 0.{\arcsec}333. The NVST was
also pointed to AR 12192 on October 25, and two series of H$\alpha$
and TiO images were obtained. The H$\alpha$ 6562.8 {\AA}
observations were from 04:59:20 UT to 05:59:15 UT with a cadence of
12 s, and have a FOV of 152{\arcsec} $\times$ 152{\arcsec} with a
spatial sampling of 0$\arcsec$.164 pixel$^{-1}$. We also use one TiO
7058 {\AA} image taken at 05:50:03 UT with a pixel size of
0.{\arcsec}052 to show the AR appearance. The Level 0 H$\alpha$ and
TiO images are firstly calibrated to Level 1 including dark current
subtraction and flat field correction, and then the Level 1 images
are further reconstructed to Level 1+ by speckle masking (see
Weigelt 1977; Lohmann et al. 1983).

Moreover, we also use the Helioseismic and Magnetic Imager (HMI;
Scherrer et al. 2012; Schou et al. 2012) and the Atmospheric Imaging
Assembly (AIA; Lemen et al. 2012) observations from the \emph{SDO}
during the \emph{IRIS} observational periods. We use the full-disk
HMI line-of-sight magnetograms with a spatial sampling of
0$\arcsec$.5 pixel$^{-1}$ and a cadence of 45 s. The AIA monitors
the Sun with a pixel size of 0$\arcsec$.6 and a cadence of 12 s.
Here, the AIA 171 {\AA} and 131 {\AA} images are chosen to study the
light wall at different temperatures. Then, we co-align all the
\emph{IRIS}, NVST, and \emph{SDO} images with the cross-correlation
method according to specific features (Yang et al. 2015).

\section{RESULTS}

\begin{figure*}
\centering
\includegraphics
[bb=40 177 506 653,clip,angle=0,width=0.8\textwidth] {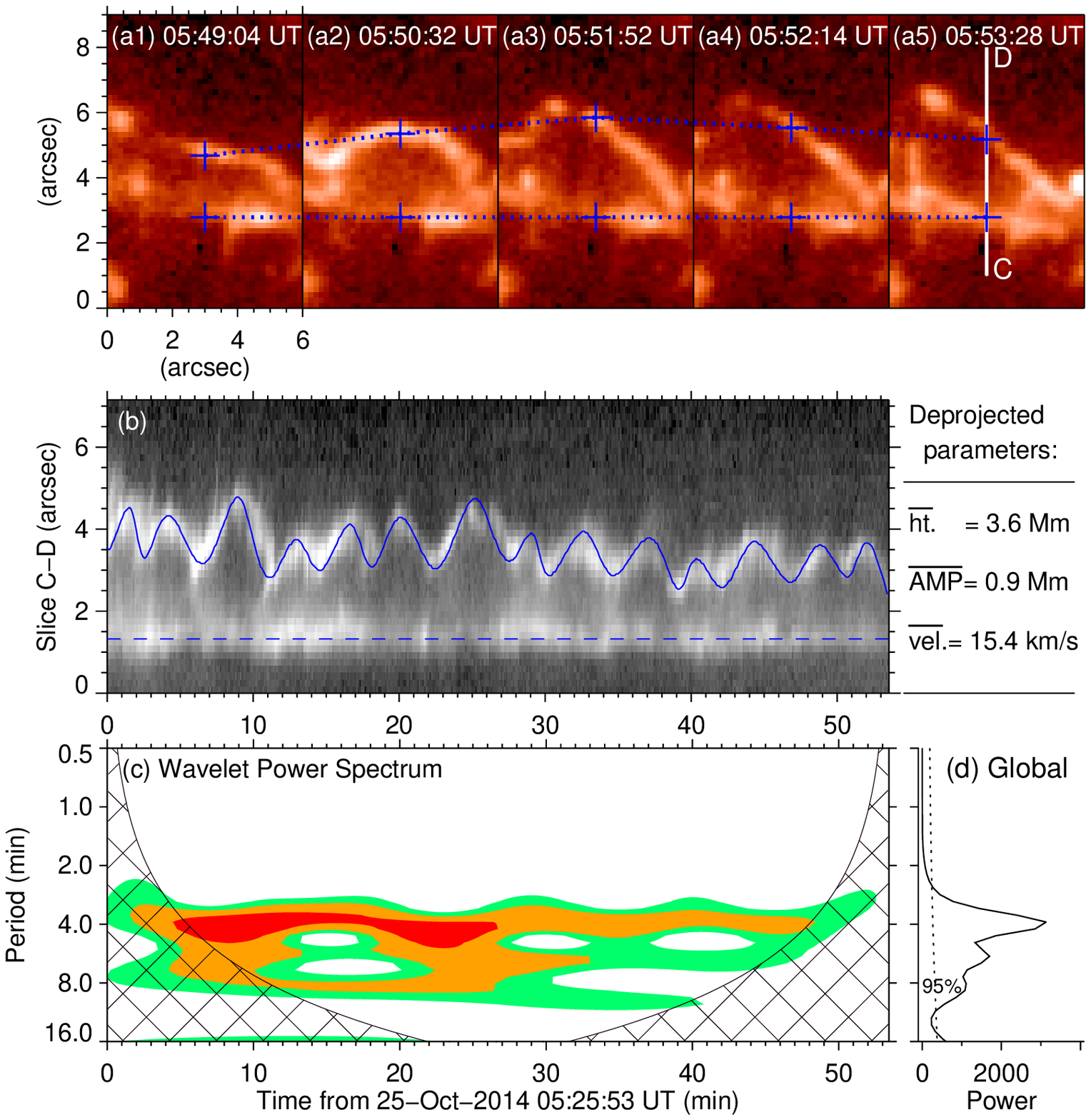}
\caption{Panels (a1)-(a5): sequence of 1330 {\AA} images showing the
evolution of a section of the light wall. The upper plus symbols and
the lower ones mark the top and base of the wall at different times.
Panel (b): space-time plot along slice ``C-D" marked in panel (a5).
The solid curve delineates the wall top, and the dashed line marks
the wall base. Panels (c) and (d): wavelet power spectrum and the
global power of the light wall oscillations. \label{fig}}
\end{figure*}

\begin{figure*}
\centering
\includegraphics
[bb=73 165 499 657,clip,angle=0,width=0.65\textwidth] {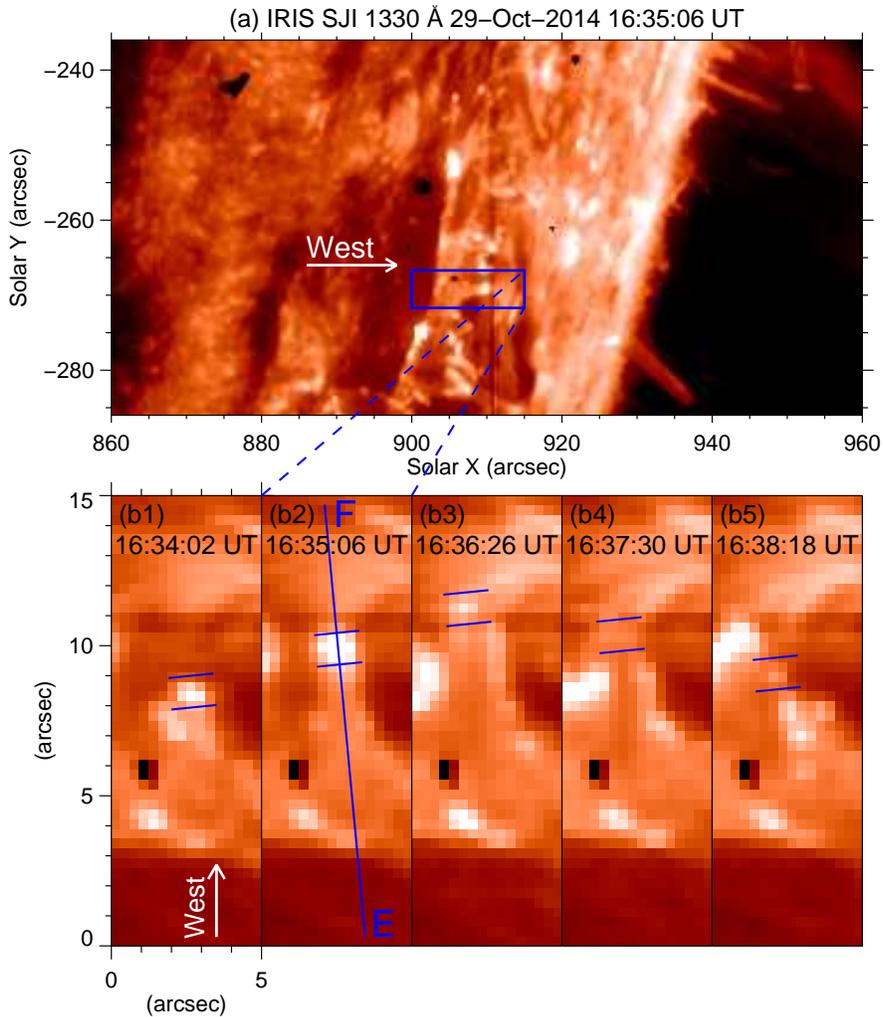}
\caption{Panel (a): \emph{IRIS} 1330 {\AA} image showing the light
wall on October 29 (also see animation 2). Panels (b1)-(b5): 1330
{\AA} images at different times showing the evolution of a section
of the light wall outlined by the rectangle in panel (a). Two short
parallel lines in each panel mark the wall top, and line ``E-F" in
panel (b2) marks the position along which the space-time plot shown
in Figure 5(a) is obtained. \label{fig}}
\end{figure*}

\begin{figure*}
\centering
\includegraphics
[bb=73 191 495 631,clip,angle=0,width=0.67\textwidth] {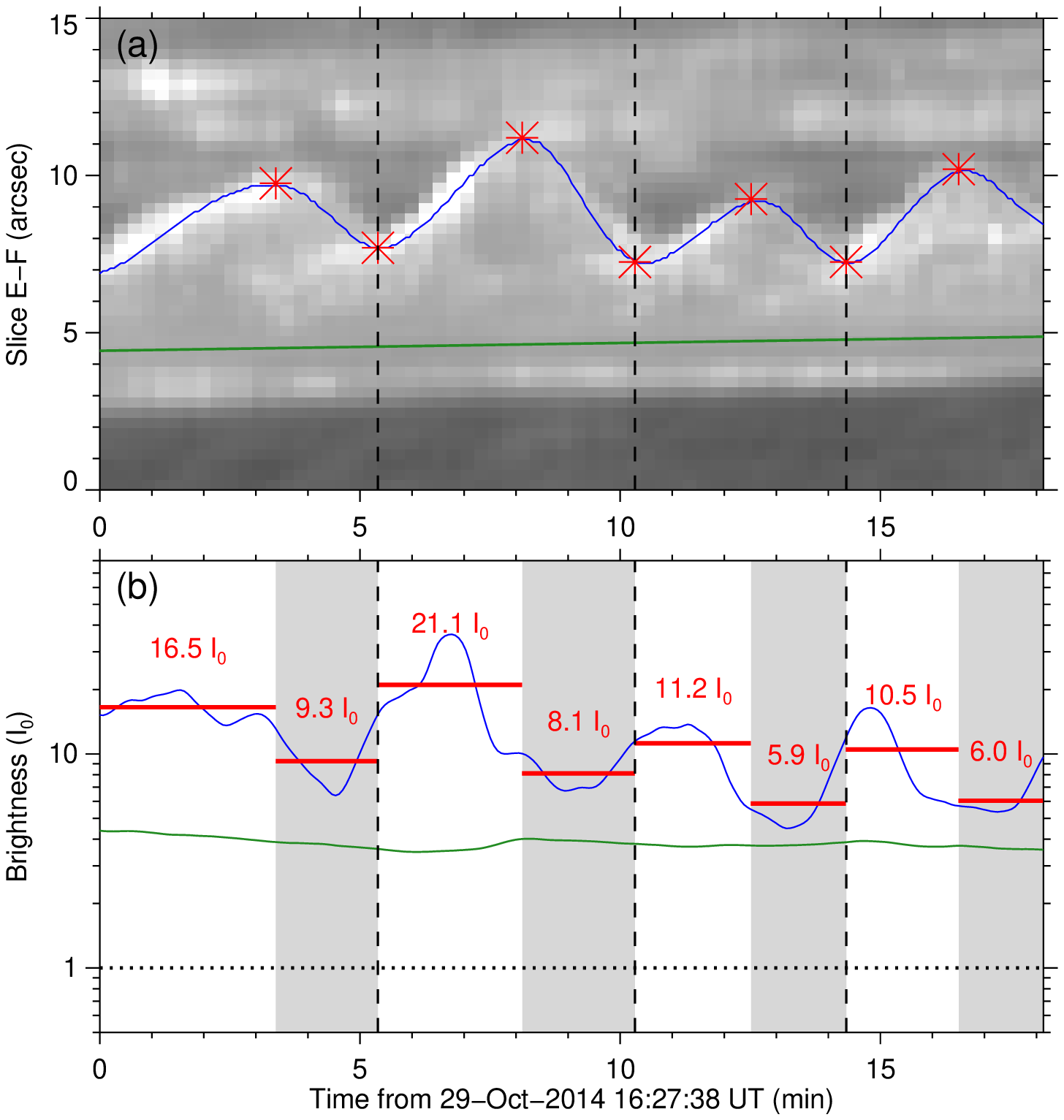}
\caption{Panel (a): space-time plot derived along slice ``E-F"
marked in Figure 4(b2). The blue curve outlines the top of the light
wall. The red asterisks mark the locations of the maxima and the
minima of four periods which are separated by three vertical dashed
lines. Panel (b): brightness of the wall top (blue curve) and wall
body (green curve) along the blue curve and green line in panel (a).
The dotted line is the brightness of the umbra. The white and gray
regions mark the upward and downward motion stages, respectively.
The red lines associated with labels indicate the average brightness
in different stages. \label{fig}}
\end{figure*}

The overview of AR 12192 on October 25 is displayed in Figure 1(a).
The polarities of the leading and following sunspots are positive
and negative, respectively. In the following sunspot, there is a
weak magnetic field channel in the strong fields (outlined by the
white square). The FOV outlined by the white square in panel (a) is
partially covered by NVST TiO observations. The TiO image (panel
(b)) which is rotated 135{\degr} anticlockwise shows the fine
structures of the umbra and penumbra. There is a distinct strong
light bridge (denoted by the arrow) crossing the umbra. The width of
the photospheric light bridge distinguished in the TiO image is
about 3{\arcsec} (2.2 Mm). The corresponding \emph{IRIS} 1330 {\AA}
image is displayed in panel (c). In the 1330 {\AA} image, many
bright structures are rooted in the light bridge, forming a light
wall (denoted by the arrow). To view the light wall well, the
expanded image outlined by the rectangle is shown in panel (d). The
light wall is brighter compared with the surrounding areas.
Especially, the top and base (indicated by the arrows) of the wall
are brighter than the wall body.

\subsection{Multi-wavelength appearance of the light wall}

We use the multi-wavelength observations to study the properties of
the light wall in detail. The NVST H$\alpha$, SJI 1330 {\AA}, AIA
171 {\AA}, and 131 {\AA} images around 05:50 UT on October 25 are
shown in Figures 2(a)-(d), respectively, and the FOV is outlined by
the square in Figure 1(d). In the 1330 {\AA} image in Figure 2(b),
there is a bubble-like structure (delineated by the black curve)
between the top (upper bright structure) and base (lower bright
structure) of the light wall. We can see that the bubble also
corresponds to a dark region in H$\alpha$ (panel (a)), 171 {\AA}
(panel (c)), and 131 {\AA} (panel (d)) images. The top of the light
wall appears as a bright structure in 171 {\AA} and 131 {\AA}
images, while it could not be identified in the H$\alpha$ image. On
the contrary, the base of the light wall appears as a bright
structure in H$\alpha$ and 1330 {\AA}, which cannot be found in 171
{\AA} and 131 {\AA} images.

We make the multi-wavelength cuts along slice ``A-B" (see the white
line in Figure 2(b)) and present them in Figure 2(e). The deep,
light, and deep pink shadows from left to right mark the locations
of the base, body, and top of the light wall, respectively. In the
H$\alpha$ cut (the green curve), the high emission corresponds to
the wall base location. The further the location away from the base
is, the lower the H$\alpha$ emission is. In the 1330 {\AA} cut (the
dark curve), there are two peaks. The larger one (marked with the
cross symbol) corresponds to the top of the light wall, and the
smaller peak corresponds to the wall base. The emission of the wall
body is lower than those of the wall top and base, however, it is
higher than that of the nearby area (see the dashed line). In the
171 {\AA} cut (the red curve), the high emission is at the wall top
location, and the emissions at the foot and body locations are quite
low.

\subsection{Oscillations of the light wall}

Figures 3(a1)-(a5) display the evolution of the light wall which is
outlined by the rectangle in Figure 2(b). We can see that the wall
top rose first and then fell down. From 05:49:04 UT to 05:51:52 UT,
the projected height of the wall increased by 0.8 Mm in about 3 min.
Then it decreased by 0.4 Mm in the following 1.5 min. Along slice
``C-D" marked in Figure 3(a5), we make a space-time plot, which is
displayed in panel (b). The most conspicuous behavior revealed by
the plot is that there exists continuous oscillations of the bright
top of the light wall. The solid blue curve outlines the position
evolution of the wall top, and the dashed line marks the base of the
light wall. The distance between the wall base and the wall top is
calculated as the height of the light wall. It should be noted that,
in order to reduce the effect of the wall width on the wall height,
the wall height \emph{h} is measured from the base center to the top
center. The mean projected wall height is $\overline{h} =
\frac{\sum_{0}^{n-1} h_{i}}{n}$, where $h_{i}$ is the projected
height in frame \emph{i}. The mean projected amplitude of
oscillations is $\overline{amp}= \frac{\sum_{0}^{N-1} | \Delta
H_{j}/2 |}{N}$, and the mean projected oscillation velocity is
$\overline{v}= \frac{\sum_{0}^{N-1} | \Delta H_{j}/\Delta T_{j}
|}{N}$, where $\Delta H_{j}$ and $\Delta T_{j}$ are the height
difference and time difference between the neighboring peak and
valley, respectively. The heliocentric angle ($\alpha$) of the light
wall is 26{\degr}. When considering the projection effect, the
corrected mean height of the wall is $\overline{ht.} = \overline{h}
/ \sin \alpha = 3.6 $ Mm, the mean amplitude is $\overline{AMP} =
\overline{amp} / \sin \alpha = 0.9 $ Mm, and the mean oscillation
velocity is $\overline{vel.} = \overline{v} / \sin \alpha = 15.4 $
km s$^{-1}$. We apply the Morlet wavelet method to the heights of
the light wall, and show the wavelet power spectrum in panel (c). It
clearly reveals that there is a main oscillation period around 4
min. The global power (see panel (d)) shows that the main period is
3.9 min.

On 2014 October 29, AR 12192 moved to a location close to the west
limb (Figure 4(a)). The light wall oscillated continually, and some
part of the wall top performed an interesting behavior, i.e., the
wall top was brighter in the upward motion phase than in the
downward phase. To describe this kind of performance well, a section
of the light wall (outlined by the rectangle in panel (a)) is
focused on. A series of sub-region images are displayed in panels
(b1)-(b5). At 16:34:02 UT, the top of the wall appeared as a bright
structure, as labeled between the parallel lines in panel (b1). One
minute later, the bright structure moved upward to a higher position
and became much brighter (see panel (b2)). The bright structure then
reached to the highest position, and its emission became weaker (see
panel (b3)). After that, the wall top moved downward with a weak
emission (panels (b4)-(b5)).

To further study the evolution of the light wall, we make a
space-time plot along slice ``E-F" marked in Figure 4(b2), and
present it in Figure 5(a). The blue curve in Figure 5(a) outlines
the evolution of the wall top. The wall top performed oscillations
with the average period of about 4.5 min. The red asterisks mark the
locations of the maxima and the minima of four periods. The vertical
dashed lines separate the different periods. The heliocentric angle
of the light wall is 78{\degr}. Then the deprojected mean height of
the wall is about 3.8 Mm, the mean amplitude is 1.0 Mm, and the mean
oscillation velocity is 14.4 km s$^{-1}$. These values are
consistent with those obtained using the October 25 series.
Especially, the deprojected wall heights obtained at different times
are comparable, implying that the wall width does not significantly
affect the height measurement of the light wall. It seems that, for
each period, the emission of the wall top in the upward motion stage
is higher than in the downward motion stage. To quantitatively study
this property, the brightness of the wall top is plotted with the
blue curve in panel (b). In addition, the green curve in panel (b)
is the brightness of the wall body (along the green line in panel
(a)). The mean brightness of the wall body is about 4 $I_{0}$, where
$I_{0}$ is the umbra brightness in 1330 {\AA}. We can see that the
brightness of the wall top is higher than the wall body all the
time. The mean brightness of the wall top is about 12 $I_{0}$ and
that of the wall body is about 4 $I_{0}$. The mean brightness values
in the four downward motion stages (gray shadow regions) are 9.3
$I_{0}$, 8.1 $I_{0}$, 5.9 $I_{0}$, and 6.0 $I_{0}$, smaller than
those (16.5 $I_{0}$, 21.1 $I_{0}$, 11.2 $I_{0}$, and 10.5 $I_{0}$)
in the upward stages (white regions), respectively.

\section{CONCLUSIONS AND DISCUSSION}

With the high tempo-spatial \emph{IRIS} 1330 {\AA} observations, we
find an ensemble of oscillating bright features in the chromosphere
and transition region above a light bridge, and name this ensemble
with a new term \emph{light wall}. The light wall is brighter than
the surrounding regions, and the top and base of the light wall are
much brighter than the wall body. In addition, we use the NVST and
AIA multi-wavelength data to study the properties of the light wall.
We find that the wall top appeared as a similar bright structure in
1330 {\AA}, 171 {\AA}, and 131 {\AA} images while it can not be
identified in the H$\alpha$ line. The wall body appeared as a bright
structure in 1330 {\AA} and as a dark structure in the other lines.
The wall top moved upward and downward successively, performing
continuous oscillations. The deprojected average height, amplitude,
and oscillation velocity are about 3.6 Mm, 0.9 Mm, and 15.4 km
s$^{-1}$, respectively. By applying the wavelet analysis method to
the heights of the light wall, we find that the main oscillation
period is 3.9 min. In another series of \emph{IRIS} 1330 {\AA}
images, we find that the wall top in the upward motion phase was
brighter than in the downward phase.

The previous studies (Sobotka et al. 2013; Yuan et al. 2014b) about
LB oscillations were done through examining the intensity variations
or the Doppler characters in LBs. In the present paper, we find a
dynamical light wall rooted in the LB, and have directly observed
the light wall oscillations in height for the first time. The
dominant period of the light wall oscillations is 3.9 min,
comparable with that (4.2 min) determined by Sobotka et al. (2013).
We interpret the oscillations of the light wall as the leakage of
\emph{p}-modes from below the photosphere. The \emph{p}-modes are
global resonant acoustic oscillations appearing as photospheric
velocity and intensity pulsations. The \emph{p}-modes can leak
enough energy to drive upward flows (De Pontieu et al. 2004),
resulting in the oscillations of the light wall.

Over the light bridge in NOAA AR 10132, Berger \& Berdyugina (2003)
observed persistent brightness enhancement in the 1600 {\AA} images
from the \emph{Transition Region and Coronal Explorer}
(\emph{TRACE}) satellite. They interpreted the enhanced LB
brightness in 1600 {\AA} as magnetic heating through some kind of
magnetic reconnection which is still unclear. According to Klimchuk
(2006), oscillations can carry acoustic flux, a significant energy
flux, to higher atmosphere to heat the solar atmosphere. Moreover,
in a study of the LB in NOAA 11005, Sobotka et al. (2013) measured
the acoustic power flux leaking along the magnetic fields in the LB
from the photosphere to the chromosphere, and found that the
transferred energy flux is sufficient to balance the total radiative
losses of the LB in the chromosphere. In the present study, for the
light wall on October 25, the wall top appeared as a constant
enhancement in brightness, which can be observed in 1330 {\AA}, 171
{\AA}, and 131 {\AA} lines, implying that there exists some kind of
atmospheric heating. However, it is difficult to know which one is
the exact mechanism for the wall top heating: the persistent
small-scale reconnection or the magneto-acoustic waves. For the
light wall on October 29, the wall top was significantly brighter in
the upward motion stage than in the downward phase, implying that
there was an intermittent heating during each oscillation period. We
suggest that this kind of dynamical behavior may be powered by the
energy released due to intermittent impulsive magnetic reconnection.

\acknowledgments { We thank the referee for his/her valuable
suggestions. This work is supported by the National Natural Science
Foundations of China (11203037, 11221063, 11303049, and 11373004),
the CAS Project KJCX2-EW-T07, the National Basic Research Program of
China under grant 2011CB811403, and the Strategic Priority Research
Program$-$The Emergence of Cosmological Structures of the Chinese
Academy of Sciences (No. XDB09000000). The data are used by courtesy
of \emph{IRIS}, NVST, and \emph{SDO} science teams. \emph{IRIS} is a
NASA small explorer mission developed and operated by LMSAL with
mission operations executed at NASA Ames Research center and major
contributions to downlink communications funded by the Norwegian
Space Center (NSC, Norway) through an ESA PRODEX contract. }

{}

\end{document}